\documentclass[11pt]{article}
\usepackage{moriond,epsfig}
\usepackage[super,sort&compress]{natbib}
\usepackage{amsmath,amssymb}
\usepackage{graphicx}

\def\MyPhoto{\pdfimage width 35mm {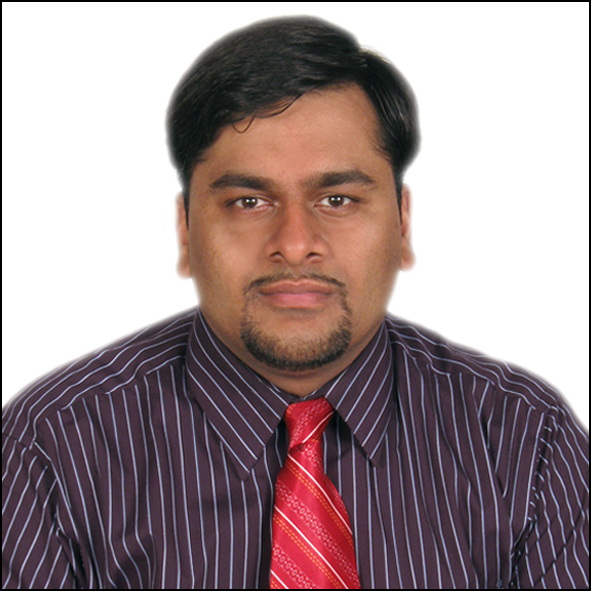}}

\def\be{\begin{equation}}
\def\ee{\end{equation}}
\def\bea{\begin{eqnarray}}
\def\eea{\end{eqnarray}}

\newcommand{\hfn}{{\sc hfn}}
\newcommand{\hfa}{{\sc hfa}}
\newcommand{\dar}{{\sc dar}}
\newcommand{\hfndar}{{\sc dar+hfn}}
\newcommand{\hfnhfa}{{\sc hfa+hfn}}

\begin{document}

\newcommand{\mnv}{\mathrm{MINER}{\nu}\mathrm{A}}

\vspace*{4cm}
\title{New approach to anti-neutrino from muon decay at rest}

\author{ Sanjib Kumar Agarwalla~\footnote{Invited talk in the Electroweak session of the Rencontres de Moriond, 2011, La Thuile, Italy.}}

\address{Instituto de F\'{\i}sica Corpuscular, CSIC-Universitat de Val\`encia, \\
Apartado de Correos 22085, E-46071 Valencia, Spain}

\maketitle\abstracts{Neutrino physics is going through a very exciting phase. 
In last one and half years, crucial informations have been provided by both short and
long baseline neutrino oscillation experiments. At short-baseline, recent neutrino 
oscillation studies seem to point towards the existence of active-sterile mixing. 
On the other hand at long-basline, recent T2K and MINOS data are in favor of non-zero 
$\theta_{13}$ opening up the possibility of observing CP-violation in the lepton sector.     
A stopped pion source provides neutrino beams with energy of a few tens
of MeV from pion and muon decay-at-rest. A rich physics program can be accomplished
with such a neutrino source. We discuss the role of such a neutrino facility to test
short-baseline anomalies and to study CP violation in active neutrinos.
}

\section{Introduction}
\label{sec:intro}

Neutrino physics is now all set to move into the precision regime, with the emphasis now shifting
to detailed knowledge of the structure of the neutrino mass matrix, accurate reconstruction
of which would unravel the underlying new physics that gives rise to neutrino mass and
mixing. In last couple of years, we are blessed with fantastic data which have been provided by 
both short and long baseline neutrino oscillation experiments. 

Recent results from short-baseline (SBL) neutrino oscillation studies seem to point towards the existence of active-sterile mixing.     
The MiniBooNE experiment has reported an apparent excess of $\bar\nu_e$ events in a beam of $\bar\nu_\mu$ 
above $475\,\mathrm{MeV}$~\cite{AguilarArevalo:2010wv} which is consistent with two-neutrino 
$\bar\nu_{\mu} \to \bar\nu_e$ oscillations at 99.4\% confidence level. This result supports the claim of the
LSND experiment~\cite{Athanassopoulos:1995iw,Aguilar:2001ty},
which has reported a $3.8\,\sigma$ excess of $\bar\nu_e$ events in a beam of $\bar\nu_\mu$.
If one interprets these results with neutrino oscillation the relevant parameter is the ratio of the
distance $L$ to the neutrino energy $E$, the so called $L/E$. The $L/E$ ratio is indeed very similar between LSND and MiniBooNE. 
The oscillation interpretation of LSND and MiniBooNE points to a mass squared difference of the order $0.1-10\,\mathrm{eV}^2$ 
and hence requires a sterile neutrino. More motivation has been provoked from a recent reanalysis of the expected
$\bar \nu_e$ flux emitted from nuclear reactors~\cite{Mueller:2011nm} that leads to an observed deficit of $\bar{\nu}_e$ at 98.6\% C.L.. 
The overall reduction in predicted flux compared to the existing data from SBL neutrino
experiments can be interpreted as oscillations at baselines of order
10--100~m~\cite{Mention:2011rk} consistent with the LSND and MiniBooNE
anti-neutrino results. 

In the month of June, 2011, new exciting results have been announced by the T2K and MINOS long-baseline (LBL) accelerator neutrino
oscillation experiments which are sensitive to $\theta_{13}$ driven $\nu_{\mu} \to \nu_e$ appearance channel.   
The T2K experiment in Japan has reported an indication of electron neutrino appearance from an accelerator-produced
off-axis muon neutrino beam of energy about 0.6 GeV produced at J-PARC~\cite{Abe:2011sj}. 
They have observed six electron-like events with an estimated background of 1.5 events in the
Super-Kamiokande detector at a distance of 295 km from the J-PARC which indicates   
towards a non-zero value of $\theta_{13}$ at $2.5\,\sigma$ significance. Within a couple of weeks of the 
T2K results, the MINOS collaboration has announced the observation of 62 electron-like
events with an estimated background of 49 events~\cite{minos}. This favors a non-zero 
$\theta_{13}$ at $1.5\,\sigma$. A latest global fit of all the available neutrino oscillation data~\cite{Fogli:2011qn} 
indicates non-zero $\theta_{13}$ at more than $3\,\sigma$ C.L..
The results on $\theta_{13}$ from these experiments are going to play a crucial role in exploring CP violation in 
future large scale experimental program of long-baseline neutrino experiments~\cite{Bandyopadhyay:2007kx}.

The pion decay-at-rest (DAR) chain leads to a beam dominated by neutrinos between 20 and 52.8 MeV, 
with a well-defined flavor content of $\nu_e$, $\nu_\mu$ and $\bar \nu_\mu$. The source may be provided 
by a low energy proton accelerator with a beam impinging on a target/dump. Potentially, this can be the cyclotrons 
planned for the DAE$\delta$ALUS CP-violation search~\cite{Conrad:2009mh,Agarwalla:2010nn,Alonso:2010fs}.
In view of the recent SBL anomalies, we discuss in the first half of my talk to repeat the original LSND experiment 
using Super-Kamiokande, doped with Gadolinium, as detector which can be coupled with a modest-power 
DAR neutrino source~\cite{Agarwalla:2010zu} positioned within 20 m of the detector. 
Then in the second half of my talk, we present the possibility to replace 
the anti-neutrino run of a long-baseline neutrino oscillation experiment, with anti-neutrinos from muon decay at rest. 
The low energy of these neutrinos allows the use of inverse beta decay for detection in a Gadolinium-doped water 
Cerenkov detector. We show that this approach~\cite{Agarwalla:2010nn} yields a factor of five times larger anti-neutrino event sample. 
The resulting discovery reaches in $\theta_{13}$, mass hierarchy and leptonic CP violation are compared with those 
from a conventional superbeam experiment with combined neutrino and anti-neutrino running. 

\section{The Neutrino Source and Decay-at-rest Flux}

In a stopped pion source a proton beam of $\sim 1$ GeV energy interacts in a low-A target 
producing $\pi^+$ and, at a low level, $\pi^-$ mesons. The pions then are brought to rest in a 
high-A beam stop. The $\pi^-$ will be captured. The $\pi^+$ will produce the
following cascade of decays
\begin{eqnarray*}
\pi^+ &\rightarrow &\mu^+ +\nu_\mu \\
&& \hspace{0.1cm}\raisebox{0.5em}{$\mid$}\!\negthickspace\rightarrow\ e^+ + \nu_e +\bar\nu_\mu 
\end{eqnarray*}
resulting in $\nu_\mu$, $\bar\nu_\mu$ and $\nu_e$, but no $\bar\nu_e$. The resulting flux is isotropic. 
As a model of a DAR source, we use the DAE$\delta$ALUS design~\cite{Alonso:2010fs}. 
The DAE$\delta$ALUS accelerators are cyclotrons~\cite{Alonso:2010yz,Calabretta:2010mh, Calanna:2011gm}, 
an ideal low-cost source for low energy (800 MeV) protons. A detailed description of the neutrino source and DAR flux 
can be found in~\cite{Agarwalla:2011qf}.

\section{Final Verdict on LSND and MiniBooNE}

\begin{table}[]
\begin{center}
\begin{tabular}{|c|cccc|}
\hline
$\Delta m^2$ $[\mathrm{eV}^2]$&0.1&1&10&100\\
\hline
signal&29&1605&1232&1314\\
\hline
\end{tabular}
\caption{\label{tab:events} Number of signal events after one year for
  $\sin^22\theta=10^{-3}$ including efficiency and energy resolution. }
\end{center}  
\end{table}

We suggest to perform a modern version of LSND, {\it i.e.} use $\bar\nu_\mu$ from a stopped pion source and 
inverse beta decay to detect the appearance of $\bar\nu_e$. The main difference
with respect to the original LSND experiment is that we suggest to use Super-Kamiokande doped with Gadolinium 
as detector~\cite{Beacom:2003nk} instead of a liquid scintillator detector. 
Super-Kamiokande has a fiducial mass of $22.5\,\mathrm{kt}$ compared to around $120\,\mathrm{t}$ in
LSND. Gadolinium doping allows to efficiently detect the capture of the neutron which is produced in 
inverse beta decay with an efficiency of  67\%~\cite{Watanabe:2008ru, Dazeley:2008xk}.
Furthermore, we use an energy resolution as given in reference~\cite{:2008zn} and an energy threshold of
$20\,\mathrm{MeV}$. We consider a 100 kW average power proton cyclotron which provides 
$4\times 10^{21}$ $\bar \nu_\mu$ per year at the source. The contamination with 
$\bar\nu_e$ from $\pi^-$ decays is very small and we take a value of $4\times10^{-4}$. 
The neutrino source will be located on the axis of the cylinder which describes the
fiducial volume and will be $20\,\mathrm{m}$ away from the first
cylinder surface. The resulting signal event rates for one year of
operation are shown in table~\ref{tab:events} and the background event
rate due to beam contamination is $765$. 

\begin{figure}[tp]
\begin{center}
\includegraphics[width=0.49\textwidth]{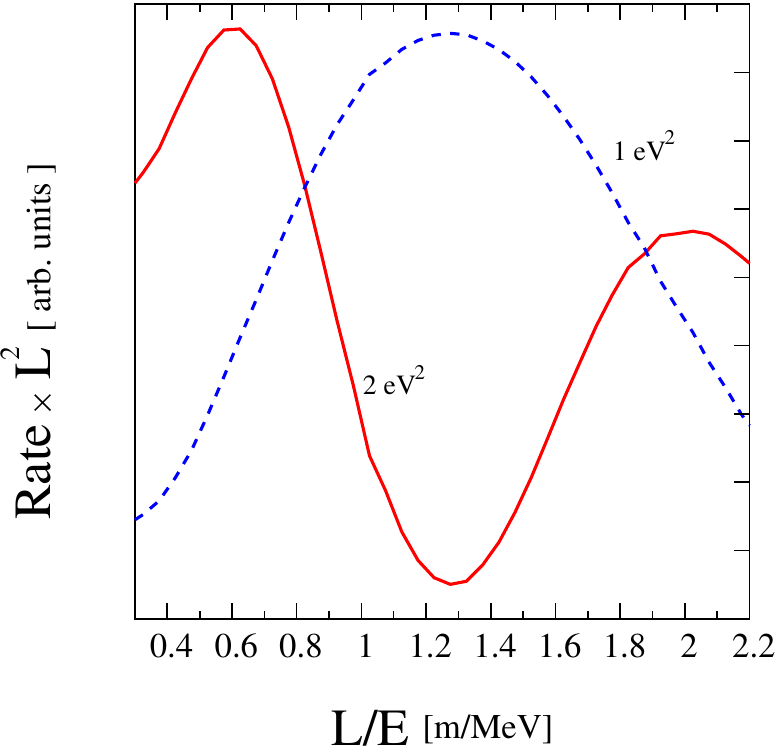}
\includegraphics[width=0.49\textwidth]{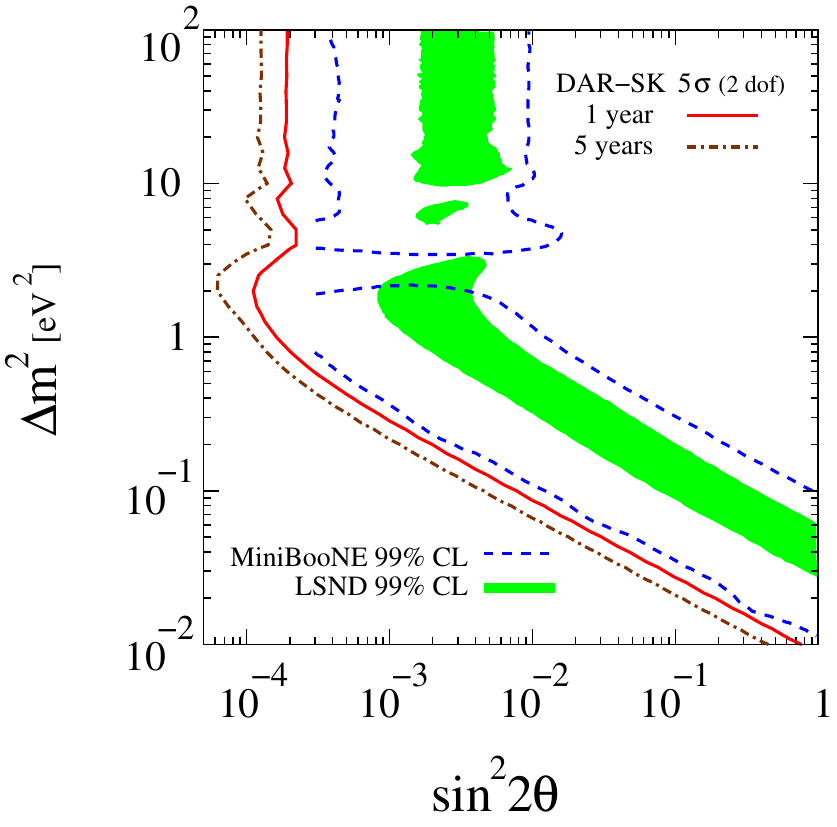}
\caption{\label{fig:spec-sens} Left panel shows the signal event rate after one year weighted with $L^2$ 
as a function of the reconstructed $L/E$. The oscillation signal is computed for $\sin^22\theta=10^{-3}$ and 
$\Delta m^2=2\,\mathrm{eV}^2$ (soild red line) and $1\,\mathrm{eV}^2$ (dashed blue line). Right panel 
depicts sensitivity limit of DAR-SK setup to sterile neutrino oscillation in the (3+1) model at $5\,\sigma$ CL (2 dof) 
using appearance mode. The solid red line corresponds to one year run of a 100 kW machine which 
can deliver $4 \times 10^{21}$ $\bar\nu_{\mu}$. The dash-dotted brown line is for five years running of a 
100 kW machine. The green/gray shaded region is the LSND allowed region at $99\%$ confidence level, 
whereas the dashed blue line is the MiniBooNE anti-neutrino run allowed region at $99\%$ 
confidence level~\cite{AguilarArevalo:2010wv}.}
\end{center}
\end{figure}

The large rock overburden of approximately $2,700\,\mathrm{mwe}$ at Super-Kamiokande, compared to 
$120\,\mathrm{mwe}$ in LSND, reduces cosmic ray induced backgrounds to negligible
levels~\cite{Conrad:2009mh,Alonso:2010fs}. Also, atmospheric neutrino backgrounds are small 
compared to the beam induced backgrounds. The large dimensions of the Super-Kamiokande fiducial volume, 
a cylinder of $14\,\mathrm{m}$ radius with a height of $36\,\mathrm{m}$ allows to
observe the characteristic baseline dependence of oscillation with
great accuracy. The size of the copper beam stop used in LSND was
about $50\,\mathrm{cm}$~\cite{Athanassopoulos:1996ds} and the position
resolution for electrons (or positrons) in Super-Kamiokande at
energies above $10\,\mathrm{MeV}$ has been measured to be better than
$75\,\mathrm{cm}$~\cite{Nakahata:1998pz}. Adding these two sources of
baseline uncertainty in quadrature we obtain about $0.9\,\mathrm{m}$.
In our analysis we account for this uncertainty by using a baseline resolution 
width of $1\,\mathrm{m}$. Thus, with a source detector distance of $20\,\mathrm{m}$ and an energy range 
from $20-52.8\,\mathrm{MeV}$ the oscillation pattern can be observed for an $L/E$ range of
$0.4-2.8\,\mathrm{m}\,\mathrm{MeV}^{-1}$. This is illustrated in the left panel of 
figure~\ref{fig:spec-sens}, where we show the signal rates weighted with $L^2$ as a 
function of reconstructed $L/E$. The oscillation signal is computed 
for two different values of $\Delta m^2$ using the usual 2 flavor 
expression with $\sin^22\theta=10^{-3}$. The ability to study the $L/E$ dependence in detail is crucial 
if a signal is observed, since it will allow to establish or refute oscillation as the underlying physical mechanism.
In the right panel of figure~\ref{fig:spec-sens} we show 
sensitivity for the $L/E$ binning analysis at $5\,\sigma$ confidence level (2 degrees of freedom) as
well as the $99\%$ confidence level allowed regions obtained from LSND and the MiniBooNE anti-neutrino 
run~\cite{AguilarArevalo:2010wv}.

\section{An Ultimate Probe for Leptonic CP violation}
\label{sec:CPV}

\begin{table}[tp]
\begin{center}
\begin{tabular}{||c||c|c||c|c||}
\hline
\hline
&$\bar\nu_\mu\rightarrow\bar\nu_e$&Background&$\nu_\mu\rightarrow\nu_e$&Background\\
\hline
\hline
{\hfndar} &1194 &217 &1532 &428\\
\hline
{\hfnhfa}&231  &158 &766 &214\\
\hline
\hline
\end{tabular}
\caption{\label{tab:rates} Comparison of the signal and background event
  rates of 6 years running of {\hfndar} and {\hfnhfa}. Note, that for  {\hfndar} this is 6 years of simultaneous running of $\nu$ and $\bar\nu$,
   whereas for {\hfnhfa} this is 3 years each, run consecutively. Oscillation parameters are $\sin^22\theta_{13}=0.1$ and $\delta_\mathrm{CP}=0$ and
  normal hierarchy.}
\end{center}
\end{table}

Here the main idea is to combine a horn focused high energy $\nu_\mu$ beam ({\hfn}) with 
$\bar\nu_\mu$ from a {\dar} setup to study $\theta_{13}$, the mass hierarchy and 
leptonic CP violation. We will denote this new technique as {\hfndar}. To illustrate the strength of {\hfndar}, 
we will study a specific setup, which closely resembles the Fermilab DUSEL concept for a long
baseline experiment, currently known as LBNE. This setup has a total running time of 6 years and a 
$300\,\mathrm{kt}$ water Cerenkov detector. The entire {\hfn} part is very similar to the setup described in detail 
in~\cite{Barger:2006vy}, specifically we take the source detector distance to be $1300\,\mathrm{km}$ and use the same detector
performance. The beam delivers $6.2\times10^{20}$ protons on target per year, which for $120\,\mathrm{GeV}$ protons 
corresponds roughly to $700\,\mathrm{kW}$ of beam power. For {\dar} setup, we consider proton cyclotrons of 1 MW beam power
which can deliver $4\times10^{22}$ of $\nu_e$, $\nu_\mu$ and $\bar\nu_\mu$ per flavor per year per cyclotron. 
We use 4 of these cyclotrons with a source detector distance of $20\,\mathrm{km}$.
In the context of superbeam experiments, a CP violation measurement requires data from both 
$\nu_\mu \to \nu_e$ \emph{and} $\bar\nu_\mu \to \bar\nu_e$. However, the horn focused high energy $\bar\nu_\mu$ beam
({\hfa}) poses a number of specific challenges: the production rate for $\pi^-$, the parent of $\bar\nu_\mu$, is lower than for $\pi^+$, 
the anti-neutrino charged current cross section is lower, the background levels are 
higher\footnote{This is due to the larger contamination from wrong sign pions.}, and the systematic 
errors are expected to be larger. Overall, the event rate for anti-neutrinos is suppressed by a factor of 2-5, 
depending on the anti-neutrino energy, which is illustrated by table~\ref{tab:rates}.

\begin{figure}[tp]
\begin{center}
  \includegraphics[width=\textwidth]{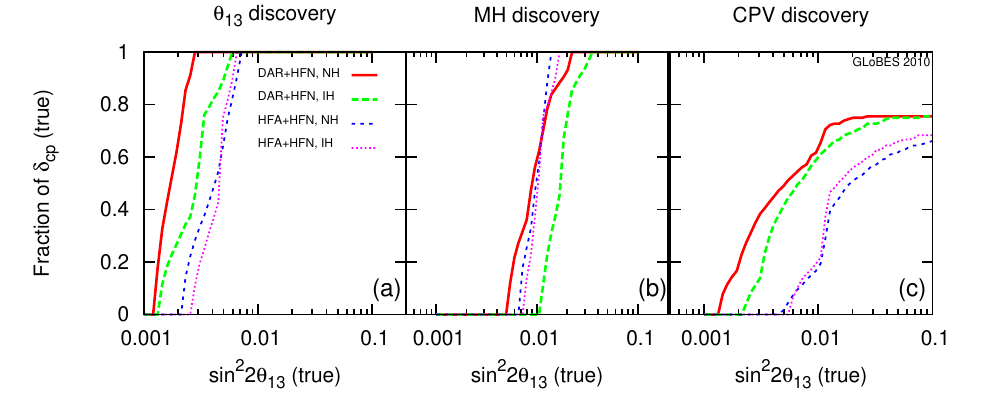}
  \caption{\label{fig:cpfrac} CP fractions for which a discovery at
    $3\,\sigma$ confidence level is possible as function of
    $\sin^22\theta_{13}$. From left to right for $\theta_{13}$, mass
    hierarchy and CP violation. The different lines are for normal
    (NH) and inverted (IH) true mass hierarchies and for {\hfndar} and
    {\hfnhfa}, respectively.}
\end{center}    
\end{figure}

In figure~\ref{fig:cpfrac}, we compare the results from {\hfndar} with
{\hfnhfa}. The reaches are given as a fraction of $\delta_\mathrm{CP}$
and as a function of the true value of $\sin^22\theta_{13}$. In panel
(a), we show the results for the discovery of the $\theta_{13}$ and
find that {\hfndar} outperforms the superbeam experiment {\hfnhfa} for
all CP phases and both hierarchies by roughly a factor two. The
discovery reach for the mass hierarchy is shown in panel (b) and here,
we see that for some values of the CP phase, in particular for
inverted mass hierarchy, the reach is somewhat smaller for {\hfndar}.
If at the end of the {\hfndar} run, the mass hierarchy has not been
discovered adding a {\hfa} run may be required. Finally, in panel (c)
the discovery reach for CP violation is shown. For $\sin^22\theta_{13}$ = 0.05,
{\hfndar} has 75\% CP coverage while {\hfnhfa} has 62\%. 

\section{Conclusions}
\label{sec:conclusions}

In this talk, we present the physics prospects of DAR neutrino sources in testing the short-baseline 
anomalies and to study CP violation in active neutrinos. We have shown that Gd doped 
Super-Kamiokande detector combined with high intensity 100 kW cyclotron DAR neutrino source  
can test the LSND and MiniBooNE claims for SBL $\bar\nu_\mu \to \bar\nu_e$ oscillations with more than 
$5\,\sigma$ significance within one year of running time. Also, we have demonstrated that a combination 
of low energy $\bar\nu_\mu$ from muon decay at rest with high energy $\nu_\mu$ from
a superbeam aimed at the same Gadolinium-doped water Cerenkov detector yields a moderately improved reach 
for $\theta_{13}$ and a significantly improved discovery reach for CP violation while only marginally
affecting the mass hierarchy sensitivity. These improvements are a direct result of combining an optimized 
neutrino with an optimized anti-neutrino run.

\section*{Acknowledgments}

I am grateful to the conference organizers for the invitation. I would like to thank J.M. Conrad, Patrick Huber, Jonathan M. Link, 
Debabrata Mohapatra and M.H. Shaevitz for their collaboration. I acknowledge the support from the European Union under the
European Commission Framework Programme~07 Design Study EUROnu, Project 212372 and the project Consolider-Ingenio CUP.

\bibliographystyle{apsrev}
\bibliography{references}

\end{document}